# Micro solid oxide fuel cells: a new generation of micro-power sources for portable applications

Francesco Chiabrera, Iñigo Garbayo, Nerea Alayo, Albert Tarancón*
Catalonia Institute for Energy Research (IREC). Jardins de les Dones de Negre 1, 2ª pl.; 08930 Sant Adrià de Besòs. Barcelona (Spain)
* atarancon@irec.cat

**ABSTRACT**

Portable electronic devices are already an indispensable part of our daily life; and their increasing number and demand for higher performance is becoming a challenge for the research community. In particular, a major concern is the way to efficiently power these energy-demanding devices, assuring long grid independency with high efficiency, sustainability and cheap production. In this context, technologies beyond Li-ion are receiving increasing attention, among which the development of micro solid oxide fuel cells (μSOFC) stands out. In particular, μSOFC provides a high energy density, high efficiency and opens the possibility to the use of different fuels, such as hydrocarbons. Yet, its high operating temperature has typically hindered its application as miniaturized portable device. Recent advances have however set a completely new range of lower operating temperatures, i.e. 350-450ºC, as compared to the typical >900ºC needed for classical bulk SOFC systems. In this work, a comprehensive review of the status of the technology is presented. The main achievements, as well as the most important challenges still pending are discussed, regarding (i.) the cell design and microfabrication, and (ii.) the integration of functional electrolyte and electrode materials. To conclude, the different strategies foreseen for a wide deployment of the technology as new portable power source are underlined.

**Keywords:** micro solid oxide fuel cell, micro power source, thin film metal oxide, ceramic, microfabrication

## 1. INTRODUCTION

Batteries, with inherent limited capacity, have dominated the power supply of small devices for decades. However, despite the fast evolution in the field, the energy gap between the capacity of the current battery technology and the power requirements is increasing year by year [1]. This energy divergence brings a great challenge on portable generation that opens new opportunities for technologies beyond Li-ion for covering the gap created in the low power regime (1 to 5W). In this new scenario, a major breakthrough on the miniaturization of uninterrupted and efficient generators is crucial.

The dream of downscaling one of the most efficient known generators, i.e. a fuel cell, has been unsuccessfully pursued for years until recent advances in the miniaturization of Solid Oxide Fuel Cells (μSOFCs) converted this disruptive technology into a serious candidate to power next generations of portable devices. Their reduced size, longer life time, high power density and possibility of integration make μSOFC very attractive. Until now, the development of μSOFC has been mainly based on the fabrication of self-supported membranes supported on microfabricated substrates [2–4]. Lately, alternative designs based on the use of porous substrates have also shown very promising results [5]. The main material used for the micromachined substrate has been silicon, due to the well-known, controlled and established microfabrication technology based on it. Other substrates have been however also utilized, including glass or metals. By creating free-standing membranes, thin electrolytes (i.e. below 500 nm) with low ionic resistance can be accessed from both sides by fuel and oxygen respectively, and operating temperatures can be lowered to < 500ºC, as compared to the typical high working temperatures of bulk SOFC systems, > 900ºC. Moreover, micro and nanotechnology has permitted reducing the size and thermal mass of the μSOFC device allowing quick and low energy consumption start-ups as well, crucial for portable applications.

Nowadays, the current technology allows operation between 350-450 ºC, with power densities as high as 1000mWcm$^{-2}$, using state-of-the-art electrolyte materials and mainly noble metals as electrodes [2,4,6–8]. However, after focusing for very long time on reducing operating temperatures, extending these range to higher temperatures (>600ºC) might paradoxically be needed for portable applications, if we want to integrate easy-to-handle and available (liquid) hydrocarbons, whose reforming typically takes place in that temperature range [9]. In this rage of intermediate temperatures, several issues related to leakage control, electrode instability and integration of µSOFC modules with other components of the complete powering system are however still to be solved.

In this manuscript, we review the status of the technology from different perspectives. First, we focus on the cell design and its integration in different supporting substrates, and second, we put the attention on the functional materials utilized for the electrolyte and the electrodes, respectively. We present as well own results on the integration in mainstream silicon technology of µSOFC devices based on free-standing large area electrolytic membranes. Full system modelling is presented for supporting the feasibility of the whole power source into real scenarios. Finally, we anticipate possible strategies for further development of the technology, towards a stable long term operation and its final deployment.

## 2. DESCRIPTION OF THE TECHNOLOGY

### 2.1 Micro SOFC concept

Microelectromechanical systems (MEMS) technology has been the main fabrication method used to produce µSOFCs. Mainstream microfabrication techniques allow large-scale and low-cost production of high quality devices. Silicon substrate-based free-standing electrolyte membranes is one of the most promising approaches for the fabrication of µSOFC power generators (PG). Among many other properties, silicon is one of the most abundant elements on the earth, and the possibility to change its electrical conductivity by simply doping or oxidizing the material makes it extremely interesting for the semiconductors field. Silicon micromechanization techniques have been developed during decades by the industry and, therefore, mass production of µSOFC power generators based on silicon substrate could be easily implemented.

Since in 2009 Evans et al. [2] reviewed the most relevant works on µSOFC devices [7,8,10–13], several have been the improvement achieved in terms of design, materials and performance. For instance, Garbayo et al. [3] developed a novel microfabrication method to obtain enlarged and more robust electrolyte membranes by using a grid of doped-silicon slabs as mechanical support. In addition, a metallic heater can be adapted to this doped-silicon nerves to instantly heat the membrane in a homogeneous mode. By means of this technology, membranes with active areas of ~8 mm$^2$ were created on a full ceramic µSOFCs that generated 100 mW/cm$^2$ power at temperatures of 750 ºC. Moreover, most of the self-supported membranes on silicon are square shaped due to the anisotropic wet etching of the substrate [7]. However, it has been demonstrated that circular shape provides enhanced mechanical stability to the membrane avoiding stresses created on the corners [14]. Thanks to the doped-silicon slabs the shape of the membrane is defined by the lithography step rather than the wet etching and circular membranes can be achieved. In Figure 1 large YSZ circular membranes supported by doped-silicon slabs can be observed. Following a similar approach, Tsuchiya et al. [4] published large YSZ membranes supported on a metallic grid. They achieved a maximum power density of 155 mW·cm$^{-2}$ at 510 ºC using a membrane with an active area of 13.5 mm$^2$. Another very interesting microfabrication approach is that suggested by An et al. [15], who in fact achieved the highest power output ever published on a µSOFC, 1.3 W·cm$^{-2}$. They enhanced the active area of the membrane by using a three-dimensional nanostructured membrane fabricated by means of nanosphere lithography (NSL) and atomic layer deposition (ALD).

Alternatives to silicon as support for the free-standing membranes can be also found in literature. Ulrich et al. [11] used Foturan, a photostructurable glass ceramic substrate. Here, the areas to be etched are exposed to UV light and crystallized at 500-600ºC. Then, the crystallized areas are selectively etched in HF. Foturan was selected as substrate material for µSOFC due to its thermal expansion coefficient (8.6x10$^{-6}$ K$^{-1}$ in the glassy state and 10.5x10$^{-6}$ K$^{-1}$ in the crystalline state), which matches with the most typical µSOFC materials.

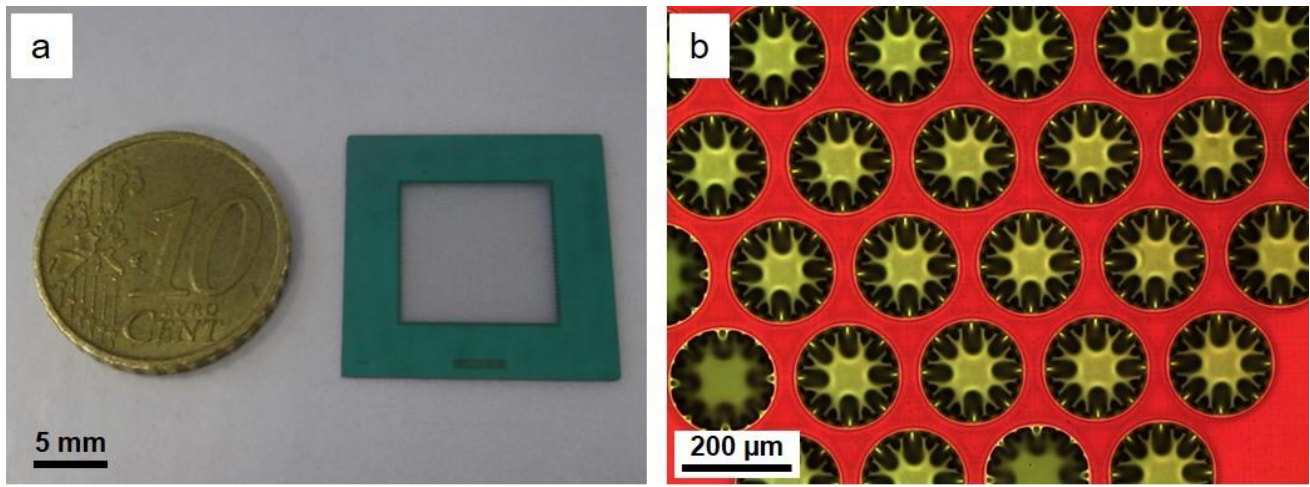

Figure 1. (a) Silicon chip with 1 cm2 membrane supported by doped-silicon slabs. (b) Microscope image of the circular YSZ membranes interconnected with the doped-silicon slabs.

Parallel to all that, several publications have appeared on the fabrication of μSOFC following a different approach, i.e. using porous supports for the thin film active layers. Joo et al. [16] proposed an anode supported μSOFC based on porous nickel thin film, avoiding lithography and etching processes. NiO paste was screen printed on a ceramic substrate and then reduced in $H_2$ atmosphere. The porous Ni film was transferred to a Ni plate and the subsequent fabrication of the fuel cell was performed on top of it. Recently, porous AAO (anodic aluminum oxide) [5] and porous stainless steel (STS) [17] anode supported μSOFC have also been presented. Porous substrates allow the simple fabrication of large area μSOFCs, as well as, potentially reduce the amount of pinholes generated due to particle injection (see section 2.2). However, the main challenge of using porous substrates is to obtain thin and high quality electrolyte throughout the entire surface due to the intrinsic inhomogeneity of the supporting material.

**2.2 The electrolyte**

The electrolyte represents the core element of thin film-based μSOFCs. It has to fulfil the following basic requirements: (i.) conduct oxide ions with low enough resistance (an Area Specific Resistance, ASR, of 0.15 $\Omega cm^2$ is generally targeted [18]), (ii.) be dense and gas tight to avoid leakages between oxidizing and reducing chambers, as well as electronic shortcuts between electrodes and (iii.) be thermomechanically stable along the whole range of operating temperatures, i.e. up to 750ºC.

The most studied and utilized materials for μSOFC electrolytes have been Yttria-stabilized Zirconia (YSZ) and Gadolinia-doped Ceria (GDC) [19,20]. Both materials have been extensively used in bulk *macro* SOFC systems, thus their mechanical and electrochemical properties are well known. While YSZ presents slightly lower ionic conductivity as compared to GDC, the main issue with GDC is its low stability under reducing atmospheres (ceria is reduced, becoming electronic conductor as well) [21]. The use of GDC is therefore usually hindered as single electrolyte layer, since it cannot be exposed to $H_2$ fuel in the anode, and typically has to be complemented with a thinner YSZ protective layer. Classically, high temperatures were needed for reaching low enough resistivity. However, by downscaling them to thin films (i.e. sub μm thickness), the operating temperature could be significantly reduced [6,22]. Great efforts have been put in the last decades on the deposition of thin electrolyte layers [20] and several approaches have been pursued, including spray pyrolysis [23–25], sputtering [26,27], Pulsed Laser Deposition (PLD) [6,19,28–31], Chemical Vapour Deposition (CVD) [32,33] or Atomic Layer Deposition (ALD) [10,34,35]. In particular, physical vapour deposition techniques, mainly PLD, ALD and sputtering, have been proven to be very effective on the deposition of dense and homogeneous layers of such complex oxide films, see e.g. Figure 2c. Importantly, they can also be adapted for large area deposition and thus easily integrated in a silicon microfabrication process for batch-production.

Depending on the μSOFC design, the challenges encountered for thin film electrolyte fabrication are different. Thus, the main challenge for electrolytes in silicon-based free-standing μSOFCs is the appearance of pinholes, mainly due to dust and/or particle ejection during thin film growth. These unwanted particles can act as shadow masks and create holes through the thin film. On the other side, the main problematic of electrolytes on the porous anode-supported μSOFCs is the ability of growing thin dense and continuous electrolyte layers on top of a highly porous supports. Figure 2a,b shows cross-section and top view optical images of a free-standing YSZ electrolyte membrane integrated in silicon. Figure 2c shows a cross-sectional exemplifying image of a free-standing YSZ layer, with platinum layers in both sides. The buckling pattern seen in Figure 2b is typical, and comes from the high compressive stress of the YSZ film [32]. This compressive stress is indeed beneficial, in order to withstand the volume changes during heating and cooling cycles. As mentioned before, the fact of being free-standing and extremely thin makes very important to control the substrate surface quality and the deposition process in order to avoid any defect on the membrane (in the form of pinholes or cracks). Figure 2d,e exemplifies the appearance of pinholes and cracks on an YSZ membrane, what provokes dramatic failure of the cells by shorting the two electrodes. The main strategies for increasing the survival rate of the membranes are (i.) a conscientious pre-cleaning of the substrates to avoid shadowing effects during deposition by the dust particles and (ii.) the use of low-grain size and thermomechanically stable targets to reduce the particle ejection while grain growth. By doing so, the density of shadowing particles could be minimized and virtually eliminated [36].

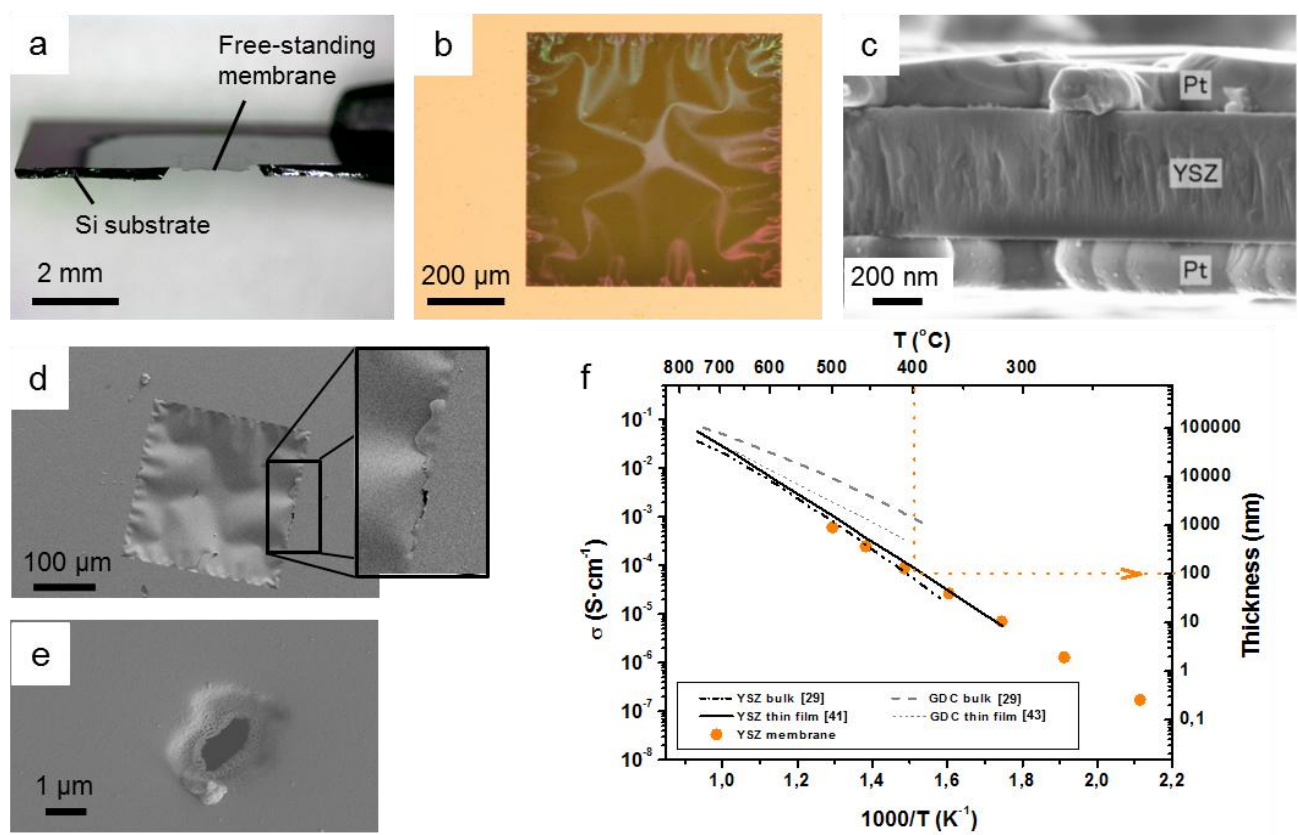

Figure 2. (a) Cross-sectional photograph of a free-standing YSZ membrane supported on a silicon platform. (b) Top view optical image of a free-standing YSZ membrane. (c) Cross-sectional SEM image of an YSZ free-standing film with Pt electrodes on both sides. (d) Top view SEM images exemplifying the appearance of cracks on a free-standing membrane. (e) SEM image of a pinhole created from a particle that shadowed the electrolyte thin film deposition. (f) Arrhenius representation of YSZ ionic conductivity on a free-standing membrane, compared to GDC and YSZ bulk and thin film references [21,28,30]. The label "thickness" in the right Y axis refers to the electrolyte thickness needed for obtaining an Area Specific Resistance of 0.15 Ωcm$^2$.

Figure 2e shows an Arrhenius representation of ionic conductivity of YSZ and GDC, in bulk and thin film form. These values are compared to the ionic conductivity measured through an YSZ free-standing membrane, supported on silicon (orange circles). As it can be observed, YSZ cross-plane conductivity matches well with the expected values according to literature for both bulk and thin films. The right Y axis displays the required film thickness for reaching the target value of 0.15 $\Omega cm^2$. Film thicknesses in the order of 100 nm are required for reaching operating temperatures of 400ºC with YSZ. This marks the lower limit in temperature for the use of YSZ as free-standing electrolyte in μSOFC. If lower operating temperatures are targeted, materials with higher conductivity would be required. On the other side, the proven thermomechanical stability of the free-standing YSZ membranes up to ~750ºC [3] makes them perfectly suited for μSOFC operation at T > 400ºC.

## 2.3 The electrodes

The electrodes in a μSOFC are of primarily importance for achieving high performance of the device, especially at low and intermediate temperature. There are some requirements common to both the oxygen electrode (cathode) and the fuel electrode (anode), which have to be fulfilled in order to assure fast electrochemical reactions of the reactants [37]. Firstly, a high electronic conductivity is requested to permit a facile current collection. In the case of using a pure electronic conductor (i.e. a metal), the area active to the electrochemical reactions is typically the triple phase boundaries (tpb), namely the points of contact between the cathode, the electrolyte and the gas phase. The use of highly porous thin films increases the tpb, allowing better electrochemical performances. When a mixed ionic electronic conductor (MIEC) is used, the active area is extended to the whole electrode surface. Nevertheless, also in a MIEC a porous structure is favoured to maximize the contact area between the gas phase and the electrode. Finally, the electrodes should be chemical compatible with the other components and stable over time, in order to assure constant characteristics over the whole life time utilization.

The most used electrodes in μSOFCs (both as anode and as cathode) are Platinum thin films [2,38,39]. Pt is known to have good electrochemical activity towards the reactants and can be easily deposited by sputtering technique, evaporation or PLD. The films deposited are typically dense and a thermal treatment is necessary to obtain the desired porosity, due to a phenomenon called dewetting [40–42]. Nevertheless, the dewetting also lead to a fast degradation of the Pt electrodes, causing the risk of loss of the electrical percolation at intermediate temperature (500ºC-800ºC) [43]. The upper part of Figure 3 shows the typical evolution of a sputtering deposited Pt thin film after annealing at 750ºC for different time. This degradation mechanism is particularly important in μSOFCs, because it can cause the loss of electric contact between the electrolyte membrane and the silicon substrate, determining the failure of the device [44]. Moreover, the use of a pure electronic conductor as electrode in a μSOFC can generate electrical constrains in the nanometric thin film electrolyte. When the film thickness is of the order of the Pt particle size, the electrolyte ionic resistance does not scale linearly with the electrolyte thickness but can saturate, leading to a decreasing of the benefits of nanometric membranes [45]. The use of electroceramic MIEC electrodes can provide a solution to both these problems [3]. Porous ceramic thin films can be easily deposited by PLD at high oxygen partial pressure [19]. The lower part of Figure 3 shows the typical evolution with time at 750ºC of a Gadolinia-doped Ceria thin film. The first annealing causes an opening of the structure porosity, which is desired to increase the active surface. After this, the structure shows a stable behaviour, with no further changes due to the temperature. The use of MIEC thin films can also avoid the undesired effect of electrolyte electrical constrains, because its active surface assures a homogenous distribution of the ionic conduction.

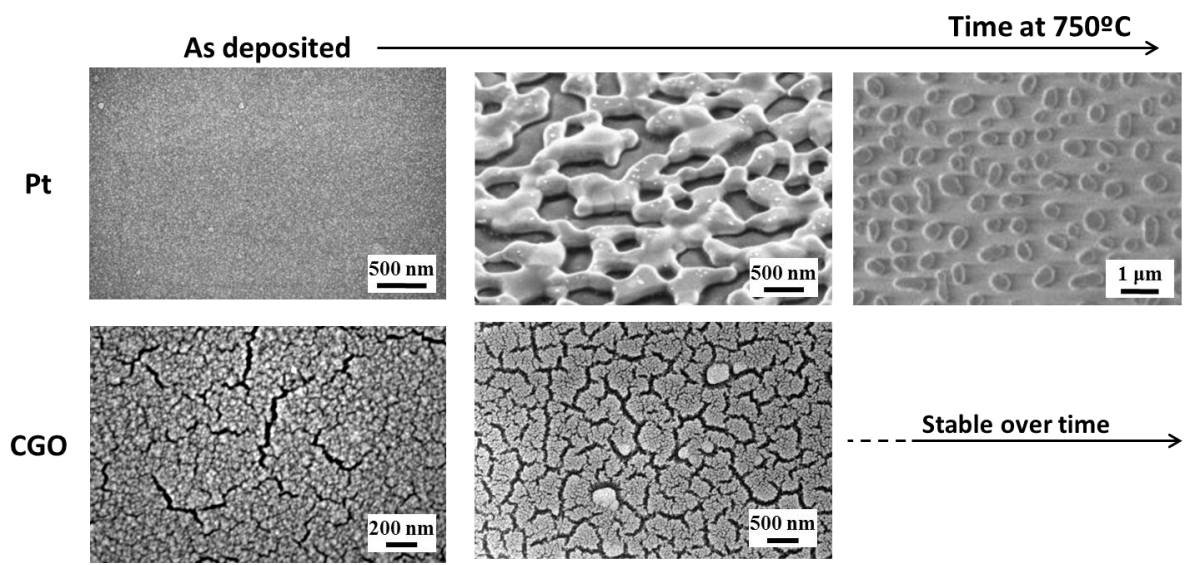

Figure 3. Typical evolution of sputtering deposited Pt and PLD Gd-doped Ceria (CGO) at high temperature. The Pt film shows dewetting and agglomeration in separated islands, while the CGO film displays, after a first enlargement of porosity, a stable behaviour over time.

Beside Pt thin films, other materials have been explored as anode in μSOFCs. The anode electrode must present a good electrochemical activity towards the oxidation of Hydrogen. In bulk SOFC, typically a cermet of Ni-YSZ is used, which combines the good catalytic activity of the Ni with the high oxygen conductivity of YSZ. A similar approach applied to anode thin films, has been used by Müller et al. [46], who synthetized by templated sol–gel chemistry coupled with dip-coating process, thin films of $NiO/Ce_{0.9}Gd_{0.1}$-$O_{2-\delta}$. Buyukaksoy and co-workers deposited thin films of NiO-YSZ, and demonstrated the feasibility of employing them as anode in μSOFCs operating at 550ºC [47]. In addition to the metal-oxides cermet, ceramic electrodes have been also explored. Porous nano-columnar $Ce_{0.8}Sm_{0.2}O_{1.9-\delta}$ have been deposited by PLD on YSZ single crystal by Jung and co-workers [48]. The MIEC thin film obtained showed low activation losses for the Hydrogen electro-oxidation due to its low tortuosity and high activity of the Ceria. Garbayo et al. deposited porous $Ce_{0.8}Gd_{0.2}$-$O_{1.9-\delta}$ by PLD and tested the anode properties in a symmetrical μSOFC [3]. The anode thin film showed good electrochemical properties at 650ºC, highlighting the possibilities of using doped ceria as single material anode in μSOFCs.

The use of cathode thin films has been extensively explored in literature. The high interest in obtaining high cathode performance is because the Oxygen reduction reactions (ORR) is one of the more limiting processes in the overall behaviour of a SOFC, especially at intermediate temperatures. Therefore, the choice of cathode materials is determinant for a large employment of μSOFCs. MIEC thin films of different chemical composition and deposited by different techniques have been investigated and tested. Thin films of $La_{0.6}Sr_{0.4}Co_{0.8}Fe_{0.2}O_3$ (LSCF) have been deposited by Ramanathan and co-workers by sputtering technique [44,49]. In their work, they measured a complete μSOFC with LSCF as cathode and Pt as anode, founding a maximum power of 60mW/cm$^2$ at 500ºC. Garbayo and co-workers deposited porous thin films of $La_{0.6}Sr_{0.4}CoO_{3-\delta}$ (LSC) by PLD [50]. They tested the electrochemical properties in a symmetrical μSOFC, with YSZ as electrolyte, finding values of area specific resistances lower of 0.3 Ω/cm$^2$ at 700ºC. Moreover, they demonstrated that the porous film was stable at this temperature and tested a fully ceramic μSOFC for high temperature applications [3]. LSC was also deposited by salt-assisted spray pyrolysis by Evans et al. [51], proving the possibility of integrating different techniques in the creation of a complete μSOFC.

The use of MIEC ceramic electrodes appears to be the best solution for improving the performances of μSOFCs and guaranteeing a long life to the device [45]. Still, further work is necessary to study the characteristics of these promising thin films.

## 2.4 μSOFC as power generator device

The fuel cell stack is the essential element of a μSOFC based Power Generator (PG). Nevertheless, other components are also required to create a complete portable power source, such as a fuel pre-processor unit (FPU) for hydrogen production from a hydrocarbon fuel (if necessary), a catalytic post-combustion unit (CPU) for exhaust gas processing and a heat management unit consisting of a heat exchanger (HX) and a thermal insulation (INS) that defines a hot module (HM). The μSOFC PG operation is described as: fuel and air are fed separately into the FPU. The fuel is heated (and vaporized in the case of liquid fuels) in the μvaporizer. A first fraction of fuel is chemically converted into simpler molecules in the μreformer, i.e. $H_2$, CO and $CO_2$. In parallel, air is heated and provided to the cathode side of the μSOFCs. Hydrogen and air react in the fuel cell membrane, producing electrical power and heat. The unreacted fuel mixes with the remaining air in the CPU where is completely burned. Finally, clean exhaust leaves the device.

In 2008, Bieberle-Hütter et al. [52] described for the first time the design of a complete μSOFC system for portable applications. The system consisted of a microfabricated solid-oxide fuel cell stack, a gas processing unit (fuel reformer and post-combustor) and a thermal system (a fuel and an air pre-heating unit, heat exchanger and insulation). One single modular element was sized for 2.5 W electrical energy output and it had an overall volume smaller than 65cm$^3$. The modularity of the system enabled adapting it to different power needs, by repeating this module it was possible to achieve higher power. Later, the same group studied the thermal design requirements for the system by means of global steady state and energy balances [53]. Particularly, they studied the impact of the overall electrical efficiency, the air-to-fuel ratio and the non-adiabatic heat losses on the operating temperature. Their method provided a useful tool to create a first thermal design layout. However, they concluded that, for overall electrical efficiencies of 20-40% and the wide range of possible heat-loss-to-power ratios, the thermal design configuration cannot be universally applied for 2.5 to 20 $W_{el}$ systems. To improve the method, experimentally obtained current-voltage characteristics should be considered in order to obtain the electrical efficiency; the auxiliary power consumption need to be taken in consideration; and the transient behaviour for systems that are capable of part load operation or demand a quick start-up time need to be validated. In regard to this, Poulikako's group proved that, in the case of μSOFC systems, a hybrid start-up process is faster and energetically more efficient than electrical heating alone, achieving a time reduction of 139 s by ten times higher electrical energy input [32,54].

Even though the above mentioned works depict a helpful first approach to understand the complexity of the system, a complete design of the whole μSOFC PG, including the geometry and experimental data from the individual elements, was necessary to advance in the optimization of the stacking configuration, the heat management strategy and the steady-state and transient regimes. Pla et al. [9] proposed a vertical stack of all the elements that compound the whole μSOFC power generator. The vertical stack design and parameterization of each component was based on experimental work presented in previous publications of the authors [3,6,32,36]. A three-dimensional thermo-fluidic model and finite volume analysis was employed to study the μSOFC power generator of 1$W_{el}$ output in steady and dynamic conditions fuelled with ethanol. At 450 °C, the system showed a self-sustained regime of operation for an insulation configuration based on materials with a thermal conductivity of 5 mW·m$^{-1}$·K$^{-1}$ and a thickness of 10 mm. A membrane supported in a grid of silicon slabs was proposed to improve the observed inhomogeneity of the temperature distribution inside the μSOFC. A quick transient regime from room temperature was proven by employing a hybrid electrical-chemical start-up.

In addition to the modelling of the complete μSOFC power generator device, there have been some experimental efforts toward the integrations of micro-reformers on the system. Pla et al. [55] have successfully fabricated and tested a standalone micro-reformer to the in situ generation of hydrogen from hydrocarbons. The micro-reformer consists of an array of vertically aligned through-silicon micro-channels (20k micro-channels per square centimetre) and resistive metallic heater, all embedded in a low-thermal-mass structure suspended by an insulating membrane. Ethanol conversion rates of 94% and hydrogen selectivity values of 70% were obtained when using operation conditions suitable for application in micro-solid oxide fuel cells, 750 °C and fuel flows of 0.02 ml$_L$ min$^{-1}$. Bieberle-Hütter et al. [56] also fabricated a micro-reformer to covert butane in hydrogen and use it in μSOFC applications. After some modifications in design, they combined the micro-reformer with a micro-fuel cell assembly to create a thermally self-sustained micro-power plant [38]. The system was operated using n-butane in air, which is partially oxidized in the micro-reformer at 410 °C thanks to the integrated heater. A maximum power density of 47mW/cm$^{-2}$ was achieved at 565 °C for an individual membrane.

# 3. STRATEGIES FOR LARGE MICRO SOFC DEPLOYMENT

µSOFCs have experienced great advances in the last years, what have pushed the technology to new limits that were not even imagined only a few decades ago. The miniaturization of the device and its integration in silicon technology have allowed reducing the operating temperatures by several hundreds of degrees with respect to the standard SOFC. Currently, a new operating temperature range is set for µSOFC at $300 < T < 450°C$, where the electrolyte presents low ionic resistance and the electrodes are structurally stable over time.

At first, reducing the temperature was considered crucial for facilitating encapsulation and integration of other components needed in the final system (circuiting, fluidic connections…). However, nowadays the technology finds himself in a dilemma. On one side, a limit in lowering the temperature has been found at ~300°C. This is clearly insufficient for an easy manipulation and encapsulation of the system, for which going to ca. room temperature would be required. On the other side, higher temperatures would be needed for promoting hydrocarbon reforming, if integration of easy-to-handle fuels is intended. However, while at those temperatures (ca. 700°C) the electrolyte membranes have been proven to be stable, the most widely utilized metallic electrodes are quickly degraded, what drastically limits the cell lifetime and performance. µSOFC technology is therefore nowadays in a *no-man´s land*, and the perspectives of wide deployment are limited unless significant structural changes are made. Here, we envision two opposed strategies for boosting again the technology towards new more reliable operating ranges.

The first approach consists of *lowering the operating temperature* even more, ideally to values close to room temperature. In this new range, encapsulation and fluidics would be easier, potentially made of plastics, and heat management would not limit design and performance. In order to reach this goal, main challenges are, first, developing electrolytes with higher conductivity than YSZ able to operate at lower temperatures and, second, integrating electrodes with enhanced surface catalytic activity. For both, the use of nanoionic effects, i.e. actively using local defects by e.g. grain-boundary or strain engineering, can potentially be the final solution for enhancing electrochemical properties.

An opposed second approach is foreseen as well, by abandoning the highly degrading metallic electrodes and substituting them by more thermomechanically stable ceramics. This would allow *widening the operating temperature window to higher temperatures*. In this case, new possibilities appear for the use of safe and easy-to-handle hydrocarbons as fuels [55]. These could be either directly used at the µSOFC, or chemically converted into hydrogen via a fuel processing unit within the system. However, thermal management would be an important challenge here. Given the enhanced heat transfer in micro systems (via conduction, convection and radiation), delivering those high temperatures at low electrical power devices and in a small package would require an excellent insulation and a very compact integration of all the components. Moreover, the start-up process would have to allow reaching the operation temperature of the reforming/fuel cell zones quickly and with minimum energy consumption, especially for portable applications [57]. In this sense, components with reduced thermal masses are targeted [21].

# 4. CONCLUSIONS

A review of the main advances on the development of µSOFC experienced in the last years is presented in this work. Two main cell designs are highlighted, i.e. the free-standing membrane based cells and the porous anode-supported cells. While the main advantage of the first is the integration of the system in the well known and scalable silicon technology, the second offers the possibility of fabricating large but still robust cells. Nowadays, an operating temperature window has been found at $T=300-450°C$ for stable µSOFC operation, by using thin YSZ or GDC electrolyte layers and porous metallic electrodes as anode and cathode. Operating at either lower or higher temperatures is however currently hindered, by either the electrolyte-associated resistance (too high at lower temperatures) or the electrode microstructural instability (films dewet at higher temperatures over time). Widening the operating temperature window is considered crucial for the deployment of the technology. Two strategies are anticipated. First, low temperature operating µSOFC with easy encapsulation and heat management are proposed by the development of high ionic conducting electrolytes. Second, increasing the operating temperature of µSOFC above 600°C would help to integrate microreformers, for the use of safe and available hydrocarbon fuels.

## ACKNOWLEDGEMENTS

This project has received funding from the European Research Council (ERC) under the European Union's Horizon 2020 research and innovation programme (ERC-2015-CoG, grant agreement No #681146- ULTRA-SOFC)